\documentclass[twocolumn,a4paper,showpacs,showkeys,aps,nofootinbib,floatfix]{revtex4}
\usepackage{epsfig}
\usepackage{here}
\usepackage{amsmath}
\usepackage{amssymb}

\begin{document}
\def\rL{\mbox{$pp \rightarrow p K^{+} \Lambda$}\ }
\def\rLg{\mbox{$pp \rightarrow p K^{+} \Lambda \gamma$}\ }
\def\rLgg{\mbox{$pp \rightarrow p K^{+} \Lambda \gamma \gamma$}\ }
\def\rLpi{\mbox{$pp \rightarrow p K^{+} \Lambda \pi^{0}$}\ }
\def\rS{\mbox{$pp \rightarrow n K^{+} \Sigma^+ $}\ }
\def\rSo{\mbox{$pp \rightarrow p K^{+} \Sigma^0 $}\ }
\def\C{\mbox{COSY--11}\ }

\title{Threshold hyperon production in proton--proton collisions at \C} 
\author{T.~Ro{\.z}ek$^{1,5}$} 

\author{D.~Grzonka$^1$}
\author{H.-H.~Adam$^2$}
\author{A.~Budzanowski$^3$}
\author{R.~Czy\.zykiewicz$^4$}
\author{M.~Janusz$^4$}
\author{L.~Jarczyk$^4$}
\author{B.~Kamys$^4$}
\author{A.~Khoukaz$^2$}
\author{K.~Kilian$^1$}
\author{P.~Klaja$^4$}
\author{P.~Kowina$^{1,5}$}
\author{P.~Moskal$^{1,4}$}
\author{W.~Oelert$^1$}
\author{C.~Piskor-Ignatowicz$^4$}
\author{J.~Przerwa$^4$}
\author{J.~Ritman$^1$}
\author{T.~Sefzick$^1$}
\author{M.~Siemaszko$^5$}
\author{J.~Smyrski$^4$}
\author{A.~T\"aschner$^2$}
\author{P.~Winter$^1$}
\author{M.~Wolke$^1$}
\author{P.~W{\"u}stner$^6$}
\author{Z.~Zhang$^1$}
\author{W.~Zipper$^5$}

\affiliation{$^1$Institut f{\"u}r Kernphysik, Forschungszentrum J\"{u}lich, D-52425 J\"ulich, Germany}
\affiliation{$^2$Institut f{\"u}r Kernphysik, Westf{\"a}lische Wilhelms--Universit{\"a}t,  D-48149 M{\"u}nster, Germany}
\affiliation{$^3$Institute of Nuclear Physics, PL-31-342 Cracow, Poland}
\affiliation{$^4$Institute of Physics, Jagellonian University, PL-30-059 Cracow, Poland}
\affiliation{$^5$Institute of Physics, University of Silesia, PL-40-007 Katowice, Poland}
\affiliation{$^6$Zentrallabor f{\"u}r Elektronik, Forschungszentrum J\"{u}lich, D-52425 J\"ulich, Germany}

\begin{abstract} 

$\Sigma^+$ hyperon production was measured at the \C spectrometer
via the \rS reaction at excess energies of $Q$ = 13 MeV and $Q$ = 60
MeV. These measurements continue systematic hyperon production studies
via the \rL / $\Sigma^0$ reactions where a strong decrease of the
cross section ratio near threshold was observed. In order to
verify models developed for the description of the $\Lambda$ and
$\Sigma^0$ production we have performed the measurement on the
$\Sigma^+$ hyperon and found unexpectedly that the total cross section
is by more than one order of magnitude larger than predicted by all
models investigated.

After the reconstruction of the kaon and neutron four momenta, the
$\Sigma^+$ is identified via the missing mass technique. Details of
the method and the measurement will be given and discussed in view of
theoretical models.

\end{abstract}

\keywords{strangeness, kaon, near threshold hyperon production, Sigma production, \C}

\pacs{13.60.Hb, 13.75.-n, 25.40.Ve, 28.20.-n} \maketitle



\section{Introduction}



The study of hyperon production in hadron induced multi particle
exit channels like $pp \rightarrow N K Y$ involves several
aspects. The nucleon--hyperon interaction can be extracted by
analyzing the $NY$ subsystem in the appropriate kinematical
region. Closely related to that is the issue of the reaction
mechanisms of the hyperon production which have to be clarified for an
unambiguous interpretation of the data. If the hyperon production is
due to the excitation and a subsequent decay of intermediate nucleon
resonances than information about the structure of the
relevant resonances can be extracted.

\begin{figure}[H]
\begin{center}
\psfig{file=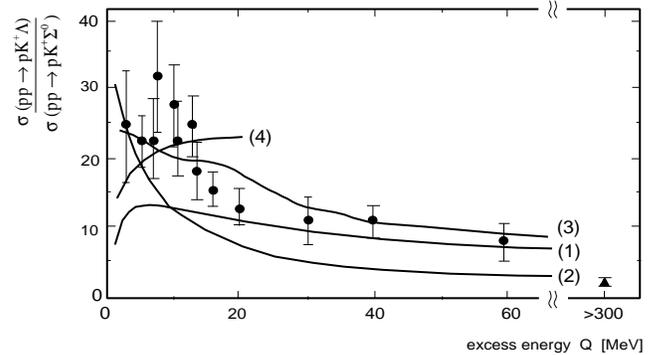,width=1.\columnwidth}
\end{center}
\vspace{-0.5cm}
\caption{\label{all_models} \small
The cross section ratio for $\Sigma^0$ and $\Lambda$ production in the
threshold region. The data~\cite{bal98,sew99,kow02,bald88} are compared
to different model predictions. The curves are theoretical predictions
described in the text.}
\end{figure}

The \rL excitation function  near threshold shows a clear deviation
from a pure s--wave phase space distribution, and the proton--hyperon final
state interaction (FSI) has to be included to describe the
data~\cite{bal98,sew99,kow02,FW}. In the \rSo channel the $pY$ FSI
seems to be negligible and the pure s--wave phase space calculations follow
the data points reasonably well. The cross section ratio $\sigma(\rL)
/\sigma(\rSo)$ below excess energies of $Q \sim$ 20~MeV is
about 28 ~\cite{sew99,kow02} in contrast to the
value of about 2.5 determined for excess energies higher than $Q =
300$ MeV \cite{bald88} (see figure~\ref{all_models}). This value is in
good agreement with the $\Lambda/\Sigma^0$ isospin relation. The
question arises if this drastic increase of the cross section ratio
near threshold is a mere effect of the $N\Lambda$ FSI or wether it
is partly due to
the reaction mechanisms in the $NY$ channels. To
explain the increase of the $\Lambda / \Sigma^0$ cross section ratio
in the  near threshold region, different scenarios were proposed.

Calculations of the strangeness production based on incoherent $\pi$ and $K$
exchange have been performed in
Ref.~\cite{sib99}. The $\pi N \to Y K$ and $K N \to Y K$ scattering
amplitudes for pion and kaon exchange, respectively, were taken from
the existing data in the higher energy region~\cite{bald88}. 
This {\it incoherent $\pi/K$ exchange model} describes the \rL cross
section over the whole energy range, however in the
near threshold region the \rSo channel is overestimated, 
and thus the predicted
$\Lambda / \Sigma^0$ ratio is too low for $Q \leqslant $ 20 MeV (see
curve (1) in figure~\ref{all_models}).

A better description of the strong rise of the ratio towards lower
$Q$-values is achieved by the {\it resonance model} (curve (2) in
figure~\ref{all_models})~\cite{sib99,sib00,shy01,tsu99}. In this model
the nonresonant direct contributions like $\pi$ or $K$ exchange were
not included, but the $\pi$, $\eta$ and $\rho$ meson exchange with the
excitation of the intermediate baryonic resonances $N$(1650),
$N$(1710), $N$(1720) and $\Delta$(1920) are taken into account. In
this {\it resonance model} the near threshold region of the
$\Lambda / \Sigma^0$ cross section ratio seems to be better reproduced
than the higher energy values \mbox{(i.e. $Q \geqslant$ 10 MeV)}. It
should be stressed that in these calculations the parameters were
fixed on the basis of higher energy data, before the
near threshold $\Lambda$ and $ \Sigma^0$ data were available.
 
Other calculations by Shyam~\cite{shy04} (based on the {\it effective
Lagrangian model}) for the strangeness production include
meson exchange ($\pi, \rho, \sigma$ and $ \omega$) together with the
excitation of resonances. The coupling constant was determined 
from data of the $\pi^+ p \to \Sigma^+ K^+$, $\pi^- p \to \Sigma^0
K^0$ and $\pi^- p \to \Sigma^- K^+$ reaction channels. The coherent
sum of resonant states and meson exchange processes describe the
experimental data for the \rL and \rSo channels very well. The {\it
effective Lagrangian model} is depicted by the curve (3) in
figure~\ref{all_models}.

The J\"ulich theory group has performed calculations including $\pi$
and $K$ exchange \cite{gas99,gas01}. In their approach the interaction
between the hyperons ($\Lambda$, $\Sigma$) and the nucleon is
described by a microscopic ($\Lambda$N-$\Sigma$N) coupled channel
model \cite{hol89} with a coherent superposition of the production
amplitudes. The $\Lambda$ production is dominated by the $K$ exchange
and therefore the contribution due to an interference between $\pi$
and $K$ exchange is negligible in this hyperon channel. On the other
hand the $\pi$ and $K$ exchanges give a comparable contribution to the
cross section in the case of $\Sigma^0$ production. An interference
between $\pi$ and $K$ exchange amplitudes act very differently on the
two channels.  Within the {\it J\"ulich meson exchange model} the
large cross section ratio can be described by a destructive
interference of the $\pi$ and $K$ exchange amplitudes only. For excess
energies above 20 MeV the model is not valid any more but
qualitatively the cross section ratio given by the model stays at a
nearly constant level. 

Although the various descriptions of the cross section ratio differ
even in the dominant reaction mechanism, all reproduce more or less
the trend of an increase of the $\Lambda / \Sigma^0$ cross section
ratio in the threshold region (see figure~\ref{all_models}). The
present data are not sufficient to 
unambiguously identify the
dominant reaction mechanism. To clarify this point
further data are needed. Especially the other isospin channels should
allow information about the production mechanisms to be extracted. Recently,
 after the
installation of a neutron detector
in addition to the $\Lambda$ and $\Sigma^0$ production the reaction channel
\rS became accessible at the \C detection system.
 The measurement of the $\Sigma^+ $
hyperon production via this reaction was performed at two beam
momenta,
\mbox{$P_{beam}$ = 2.6 GeV/c} and $P_{beam}$ = 2.74 GeV/c,
corresponding to excess energies of 13 MeV and 60 MeV,
respectively.


\section{Experiment}

\begin{figure}[H]
\begin{center}
\epsfig{file=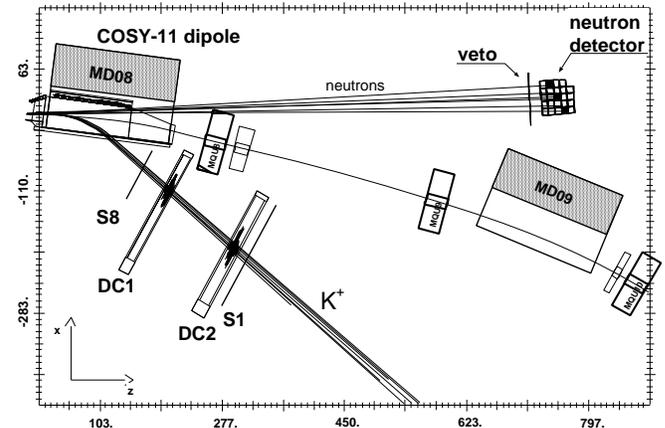,width=1.\columnwidth}
{\caption{\label{tracks}\small \C detection set-up~\cite{bra96}
with the superimposed tracks of kaons and neutrons from the
\rS \hspace{-0.2cm}. The scales of the axes are given in [cm].}}
\end{center}

\end{figure}

\C is an internal magnetic spectrometer at the COoler SYnchroton and
storage ring COSY~\cite{cosy_ring} in J\"ulich. The interaction
between a proton in the beam and a proton from the $H_2$ cluster
target~\cite{target-dom} may lead to the production of the $\Sigma^+$
hyperon in the \rS reaction. The charged reaction products are
separated from the circulating beam in the magnetic field of one of
the regular COSY dipoles~\cite{bra96}. The $\Sigma^+$ hyperon is
identified via the missing mass technique by detecting the $K^+$ and
the neutron. The momentum vector ($\vec{p}$) of the $K^+$ meson can be determined
by tracking the $K^+$ trajectory reconstructed in the drift
chambers (DC1 and DC2 in figure~\ref{tracks}) through the known
magnetic field back to the target point. The momentum vector together with the 
velocity ($\beta$)
measurement in the two scintillators S8 and S1 are used to  identify the kaon
via its invariant mass $(M_{inv}^2~=p^2(\beta^{-2}-1))$.

Assuming that a hit in the neutron detector is from a neutron, the
four momentum vector of the neutron is calculated by the measured velocity,
the direction of the neutron (given by the first module hit) and the
known mass. The background from charged particles incident on the neutron
detector is discriminated by veto scintillators.

\begin{figure}[H]
\begin{center}
\epsfig{file=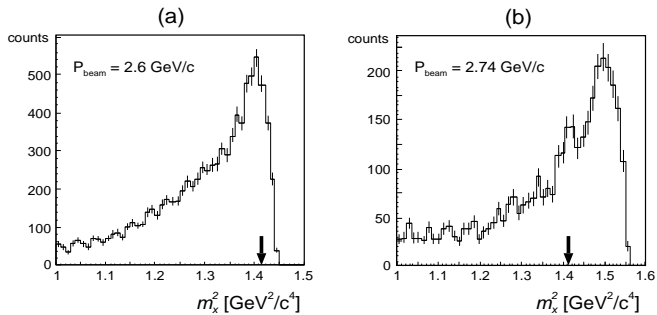,width=1.\columnwidth}
{\caption{\label{mm1-fig} \small The experimental squared missing
mass spectra of the \mbox{$pp \rightarrow n K^+ X$} system for two
investigated beam momenta. The arrows point to the nominal squared
mass of the $\Sigma^{+}$ hyperon. The statistical errors of the counting
rates are shown.}}
\end{center}

\end{figure}

In figure~\ref{mm1-fig} the experimental distributions of the squared
missing mass ($m_{x}^{2}$) of the \mbox{$pp \rightarrow n K^+ X$}
system for the two beam momenta are shown. For the higher momentum, an
enhancement around the squared $\Sigma^{+}$ mass is clearly seen on a
large background (figure~\ref{mm1-fig}b), but for the lower beam
momentum (figure~\ref{mm1-fig}a) a $\Sigma^{+}$ peak is not directly
visible. 

\begin{figure}[H]
\begin{center}
\epsfig{file=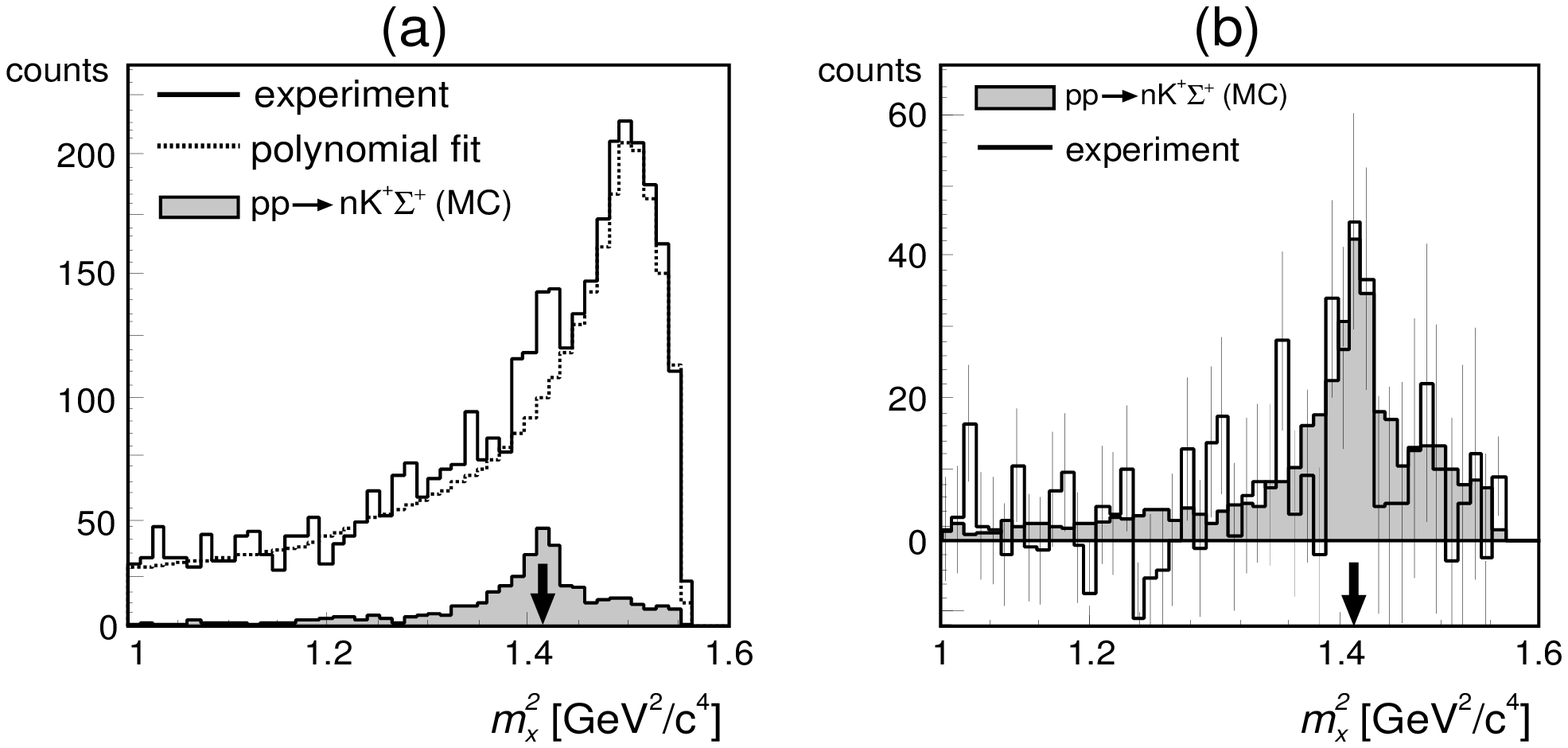,width=1.\columnwidth}
{\caption{\label{mmh_wielo-reac} \small Background determination
for the \rS reaction at \mbox{$P_{beam}$ = 2.74 GeV/c}. (a) The
experimental squared missing mass spectrum of $nK^+$ system with a
polynomial background fit and the simulated \rS
spectrum. (b) Result after subtracting the background
from the experimental distribution in comparison to the simulated
spectrum. The arrows show the nominal squared mass of the $\Sigma^+$
hyperon. Only statistical errors are shown.}}
\end{center}
\end{figure}

In order to determine the number of $\Sigma^+$ events in the higher
energy data set, 
the background has been parametrized by
a polynomial function
which is added to the expected missing mass distribution of the $nK^+$
system for the \rS reaction obtained from simulation studies. In
figure~\ref{mmh_wielo-reac}a the experimental missing mass spectrum of
the \mbox{$pp \to n K^+ X$} system is compared with the fitted
polynomial function. The expected distribution from Monte-Carlo studies with $
X = \Sigma^+$ is depicted in the figure as
well. Figure~\ref{mmh_wielo-reac}b shows the result of the subtraction
of the polynomial from the experimental missing mass
distribution together with the Monte-Carlo distribution.

In order to understand the background distribution, 22 reaction
channels (mostly multi-pion reactions but also \rL $(\Sigma^0,
\Lambda\gamma)$) were simulated and their contributions to the missing
mass distribution were determined. These studies showed that the
reactions \rL\!\!, \rLg and \rLgg are the dominant background
channels in the $\Sigma^{+}$ region. The Monte Carlo code includes the
realistic geometry and physics processes like energy loss and
straggling which occasionally cause the misidentification of the
particle type. 
 
All background channels result in a rather smooth distribution of the
missing mass spectrum as can be inferred from calculations from Monte-Carlo
studies and by comparing the two experimental distributions (see
figure~\ref{mm1-fig}). 
 
For the lower energy data set a $\Sigma^{+}$ peak is not obviously
visible via the missing mass distribution. Therefore, a simple
polynomial background parametrization cannot be used. To determine the number
of $\Sigma^{+}$ events it was assumed that the background shape for
this data set is the same as that at the higher energy. This
assumption is justified since there are no new open channels for the
higher energy. 

\begin{figure}[H]
\begin{center}

\epsfig{file=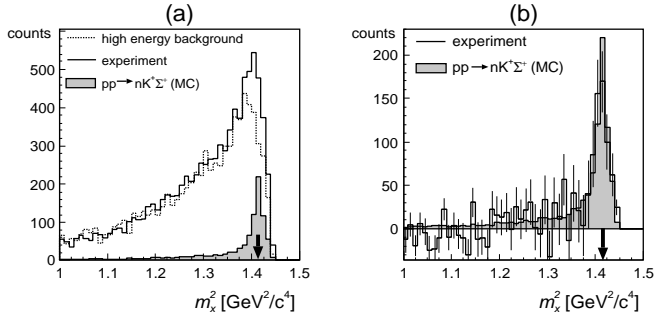,width=1.\columnwidth}
{\caption{\label{mml_fit-reac} \small Background investigation
for the \rS reaction at $P_{beam}$ = 2.6 GeV/c. (a) The experimental
squared missing mass spectrum of the \mbox{$pp \to nK^+ X$} system
is compared to the assumed background taken from the data at $Q$ = 60
MeV and the simulated \rS distribution. (b) Result of the
subtraction of the background from the experimental
distribution in comparison to the simulated spectrum. The arrows show
the nominal squared mass of the $\Sigma^+$ hyperon. Only
statistical errors are shown.}}
\end{center}
\end{figure}

At the COSY-11 experiment 
the shape of the missing mass distribution is mainly determined 
by the acceptance of the detection system and is dependent 
on the excess energy of an individual event.
From the analysis of $\eta$ and $\eta^{\prime}$ production studies at COSY-11
it was verified that the background shape 
resulting here mainly from multi pion production 
is in very good agreement
with the expectations from Monte Carlo studies 
taking into account the detector characteristics
and is comparable at different beam momenta. 
In addition, Monte Carlo data of the reaction channels which contribute
dominantly to the background in the $\Sigma ^+$ production 
were compared in view of the background shape 
by adjusting the kinematical limits. Within error bars their shapes were identical.
Therefore it is justified to assume that the background shape 
is the same for both beam momenta. 
For a detailed discussion on the background shape at COSY-11 we refer to
\cite{mos06}.
The background shape from the
experimental missing mass distribution for the higher energy data set
was shifted such that the kinematical limits for both
spectra were the same. Afterwards, it was added to the
missing mass distribution of the \rS reaction which was determined 
from Monte-Carlo studies 
at the lower energy (see
figure~\ref{mml_fit-reac}).
The relative magnitudes were determined by the $\chi^2$ minimisation
of the functional forms to the data sample at lower energy.


\section{Results}



\subsection{Total cross section}



For the lower energy data set, 
no clear enhancement near the kaon mass in the invariant mass
distribution is visible. Therefore the assumption was made, that
the kaon peak in the experimental distribution has the same position
and width as in the simulated distribution. 

\begin{figure}[H]
\begin{center}
\epsfig{file=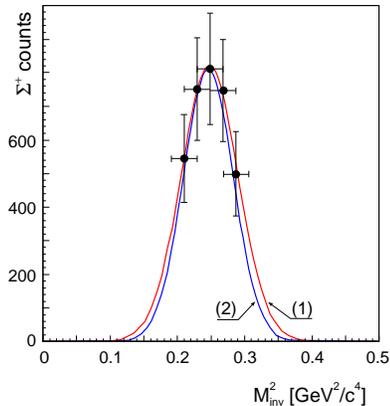,width=0.59\columnwidth}
\vspace{-0.3cm}
{\caption{\label{inv_scan} \small The number of $\Sigma^+$ events obtained for different 
intervals of the $K^+$
invariant mass. Curve (1) -- expected from the Monte-Carlo studies
distribution of the $K^+$ events. Curve (2) -- Gauss fit to the
points on the figure given by the number of $\Sigma^+$ events. See
text for details.}}
\end{center}
\end{figure}

As a cross check, event samples in different regions 
of experimental invariant mass but
still within the expected range for kaons
were taken and the missing mass distributions were determined. 
The assumed background shape was
subtracted from the experimental distribution and the enhancement
around the mass of the $\Sigma^+$ was interpreted as a signal from the
$\Sigma^+$ hyperon (see figure~\ref{mml_fit-reac}). This procedure was
repeated for five different invariant mass regions and the number of
events under the remaining peak was determined. The results are shown
in figure~\ref{inv_scan}. The vertical error bars
correspond to the statistical error of the number of events. The
horizontal error bars indicate the widths of the appropriate invariant
mass intervals.
In figure~\ref{inv_scan} curve (1) shows the expected invariant
mass distribution of kaons from Monte-Carlo calculations. Curve (2) is a Gauss fit to the
experimentally determined number of $\Sigma^+$ events in the different
invariant mass regions. This can be interpreted as an experimental
distribution of kaons. The assumed experimental kaon invariant mass
distribution is consistent with the distribution expected from Monte-Carlo
studies. 

\vspace{-0.3cm}
\vspace{-0.2cm}

For the higher beam momentum a kaon peak 
is clearly visible 
in the
invariant mass distribution 
and
can be fit by a Gauss function. In a systematical study 
of the result at the higher beam momentum 
three different selections on the experimental invariant mass
distribution were applied. Events from the following regions:
$\mu_{exp} \pm 0.25 \sigma_{exp}$, $\mu_{exp} \pm 0.5 \sigma_{exp}$
and $\mu_{exp} \pm 1.0 \sigma_{exp}$ centered around the nominal reconstructed mass ($\mu_{exp}$)
were taken and the corresponding
missing mass distributions were generated. 
$\sigma_{exp}$ is the experimental resolution of this observable.
Next the number of
$\Sigma^+$ events for each of these distributions was determined. 
The results are listed in the middle column of
table~\ref{number_higher} 
and in the last column the number of the $\Sigma^+$ events
corresponding 
to the full Gaussian distribution.

\vspace{-0.5cm}
\small
\begin{table}[H]
\begin{center}
{\caption[ The number of $\Sigma^+$ events in the higher energy data
    set.]{\label{number_higher} \small 
The number of $\Sigma^+$ events as a function of the width of the 
invariant mass range. Only statistical errors are given.}}

\begin{tabular}{|c|c|c|}
\hline 
Invariant   &Number                   & Number of events identified \\
mass range   &of events                & as $\Sigma^+$ and scaled to the \\
$\mu_{exp}$ &identified as $\Sigma^+$ & full Gaussian distribution \\

\hline
$ \pm $ 0.25 $\sigma_{exp}$ & 187 $\pm$ 51  & 944 $\pm$ 257 \\ \hline
$ \pm $ 0.5 $\sigma_{exp}$  & 367 $\pm$ 83  & 960 $\pm$ 217 \\ \hline
$ \pm $ 1.0 $\sigma_{exp}$  & 661 $\pm$ 134 & 969 $\pm$ 197 \\ \hline
\end{tabular}

\end{center}
\end{table}
\normalsize
\vspace{-0.4cm}


In order to calculate the cross section for the \rS reaction the
number of $\Sigma^+$ hyperon events and the detection efficiency of
the \C apparatus for the two excess energies were determined.
The luminosity was determined by a simultaneous measurement of proton-proton elastic scattering. 

In table~\ref{results-tab-sys} the total cross sections for both beam
momenta are given. The systematical errors are due to: 
$i)$ error of
the detection efficiency determination which is \mbox{8.5 \%} for the
lower and \mbox{3.5 \%} for the higher energy data set (including the
inaccuracy of the effective detector position and of the beam momentum
determination), 
$ii)$ uncertainty in the form of the background, and
$iii)$ error of the luminosity calculation which is
\mbox{3 \%} for both data sets and includes the uncertainty due to
the normalization procedure and the error of the solid angle
determination.
For the data at 2.74 GeV/c the uncertainty in the background form
was estimated by comparing the polynomial parametrization to
a background form resulting from the sum of known
background reaction channels generated in Monte Carlo studies. 
The difference is about 18\%.
For the data at 2.6 GeV/c the $m_{x}^2$ region 
used to normalize the
background was varied resulting in
an error of about 20\%.
The values in table~\ref{results-tab-sys} 
include also a change in
the detection efficiency resulting from the inclusion of higher
partial waves. Close to the reaction threshold 
partial wave 
contributions 
higher than $L~=~0$
are not expected. However, if the excess energy $Q$
for a given channel goes beyond a few MeV
it cannot be a priori neglected.
The contribution of partial waves with $L~>~0$ 
is not known for the reaction investigated here.
Thus, this effect has been studied by assuming a similar distribution
of partial waves as measured in the \rL channel 
studied by the TOF collaboration
at COSY~\cite{hessel,fritsch,tof_web}. 
The inclusion of  higher partial waves with a relative 
strength
quoted in~\cite{hessel,fritsch} lower the detection
efficiency by 30 \% for the lower and by 7.7 \% for the higher
energy data set. The sum of the total systematical errors
equals  \mbox{ 60 \%} for the lower and \mbox{ 34 \%} for the
higher energy data set.

\vspace{-0.7cm}
\small
\begin{table}[H]
\begin{center}  
{\caption[Values of the total cross section for the \rS
channel. Statistical and systematical errors are
presented.]{\label{results-tab-sys} \small Total
cross section for the reaction \rS.
Statistical and systematical errors are presented, respectively.}}

\begin{tabular}{|c|c|c|}
\hline 
Beam momentum & Excess energy& Total cross section \\
$P_{beam}$    & $Q$          &     $\sigma$        \\
\mbox{$[GeV/c]$}&\mbox{$[MeV]$}&      $[\mu b]$          \\
\hline
   2.60      & 13           &  $4.56 \pm 0.94 \pm 2.7$  \\ \hline
   2.74      & 60           &  $44.8 \pm 10.7 \pm 15.2$\\ \hline
\end{tabular}
\end{center}
\end{table}
\normalsize


\subsection{Comparison with model predictions}
\label{results}


Among the models described in the introduction only two give
predictions for the \rS reaction, namely the {\it J\"ulich meson
exchange model}~\cite{gas99,gas01} and the {\it resonance
model}~\cite{sib99,tsu99}.
Calculations of the $\Sigma^+$ production within the {\it J\"ulich
meson exchange model} predict a total cross section of
\mbox{$\sigma$ = 0.23 $\mu b$} at $Q$ = 13 MeV for the destructive
interference (which was necessary to describe the high
$\Lambda/\Sigma ^0$ cross section ratio at threshold). 
This is about a factor of 20 below the 
experimental value of 4.56 $\mu$b given in table
\ref{results-tab-sys}.
A constructive interference would
result in a cross section even a factor of 53 too low.

\begin{figure}
\begin{center}
\epsfig{file=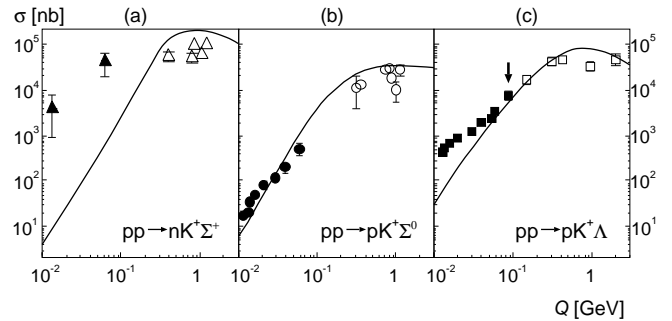,width=1.\columnwidth}
\end{center}
{\caption{\label{res-sib} \small Comparison of the experimental
total cross section with the {\it resonance
model}~\cite{sib99,tsu99} predictions for various \mbox{$pp \to N
K^+ Y$} reactions. Full triangles in (a) are data obtained in this
work. Data in the near threshold region (presented as full
symbols in (b) and (c)) are taken from
Refs.~\cite{bal98,sew99,kow02} and data from the high excess energy
region (open symbols) from Refs.~\cite{bald88,flamino}. In (c) the
data point indicated by the arrow was determined from our data as a
cross check of the luminosity calculation.}}
\end{figure}

For the {\it resonance model} the predictions for the \rS channel for
the near threshold region deviate even more from the data. In
figure~\ref{res-sib} the model predictions and the available data for
the \rS (a), \rSo (b) and \rL (c) channels are shown.
The filled  data points presented by triangles, dots
and squares in the near threshold region were measured by the \C
collaboration~\cite{bal98,sew99,kow02}. The data point in the \rL
channel indicated by the arrow was determined in parallel to this work by selecting
the $K^+ p$ exit channel which was included in the events 
sample at 2.74 GeV/c. The high energy data for the given
reactions were taken from~\cite{bald88} and~\cite{flamino}. The model
calculations for each channel are given by the solid
lines~\cite{tsu99,Sibi-pc}.

The data point for the \rS channel at \mbox{$Q$ = 13 MeV} is
underestimated in the total cross section calculated 
using the resonance model \cite{sib99,tsu99} by about a factor
of 500 and for \mbox{$Q$ = 60 MeV} by about a factor of 50. For the
\rSo channel, this  model calculation describes the existing data set
and in the case of the \rL channel the underestimation of the cross
section in the near threshold region is about a factor of 16 being
30 times smaller than for the $\Sigma^+$ production. At high excess
energies, the $\Sigma^+$ data points are by factor 3 -- 4 below
the model calculations. Previous \C hyperon production studies
conclude that final state interactions (FSI) plays an important role
in the near threshold $\Lambda$
production~\cite{bal98,sew99,kow02,bal98a}. In the {\it resonance model} the
FSI is not included~\cite{sib99,tsu99} and therefore the deviation of
the model calculations from the data points in the near threshold
region is expected if a strong FSI is present. This effect is clearly
seen for the \rL and barely observed for the \rSo reaction channel. 

\begin{figure}[H]
\begin{center}
\epsfig{file=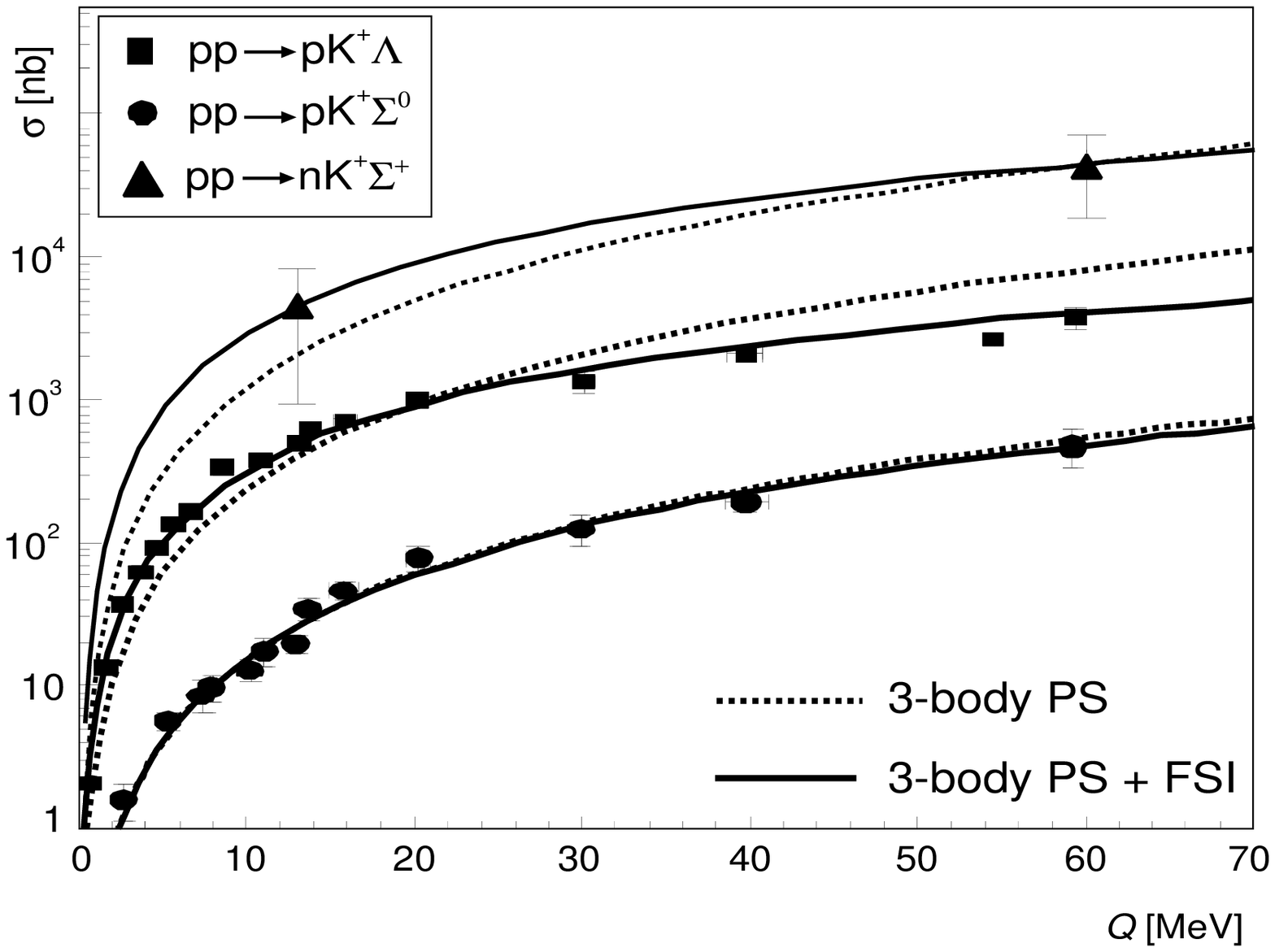,width=0.9\columnwidth}
\end{center}

  {\caption{\label{FW-all} \small The \rS, \rL and \rSo cross
  sections as a function of the excess energy $Q$. Experimental data
  are from Refs.~\cite{bal98,sew99,kow02, TOF-bilger} and from this work. 
The error bars for the \rS data represent a sum of statistical and
systematical uncertainties given in table \ref{results-tab-sys}.
The  lines show the calculations corresponding to 3-body phase space with
  (solid line) and without (dashed line) final state interaction. 
}}
\end{figure}

Investigations of hyperon production in \C~\cite{bal98,sew99,kow02}
indicate that the \rL data cannot be described by
a pure 3-body phase space (PS)
dependent cross section expressed as~\cite{byckling}:
\begin{equation}
\sigma = K \cdot Q^2,
\label{PS}
\end{equation}
where $K$ is a normalization factor and $Q$ the excess energy.
Therefore a modification is needed which
takes into account the {\it proton - hyperon} FSI. In order
to describe the near threshold region, the parametrisation of the
excitation function including the FSI proposed by
F\"{a}ldt-Wilkin~\cite{FW} was used. It is expressed by:

\begin{equation}
\sigma = C \cdot \frac{Q^2}{(1+\sqrt{1+Q/\varepsilon})^2},
\label{FW-eq}
\end{equation}
where $C$ and $\varepsilon$ are parameters related to the FSI strength.

In figure~\ref{FW-all} the cross sections for different production
channels for the $\Lambda$, $\Sigma^0$ and $\Sigma^+$ hyperons are
compared to predictions based on the 3--body phase space (PS, dotted line)
and the 3--body phase space calculations modified by the $pY$ FSI (PS
+ FSI, solid line), following equation~\ref{FW-eq} with $\epsilon$ and
$C$ as free parameters. These parameters are related to the scattering
length {\it a} and the effective range {\it r} of the $pY$ potential~\cite{FW}.

For the \rS data the resulting  $\varepsilon$ and $C$ parameters have
similar values as for the \rL channel.  It seems that for
the $\Sigma^+$ production via the \rS reaction a rather
strong $n-\Sigma^+$ FSI is present, however, within the error bars
also the curve obtained without FSI describes the two data points.
Therefore an unambiguous conclusion about the $n-\Sigma^+$ FSI  
requires more data to
disentangle the reaction mechanisms and especially the role of nuclear
resonances.


\section{Conclusions and perspectives}

        \label{perspectives}


The total cross section of the \rS reaction was determined by means of the \C
detection system for excess energies of $Q$ = 13 and 60 MeV. 
The values established are more than an order of magnitude 
larger than the expectations of currently available model predictions.

The unexpected large total $\Sigma^+$ production 
cross section is consistent with an observation 
by Tan~\cite{TAN69} who concluded
that when assuming charge symmetry in $\Sigma^+~n$ and $\Sigma^0~p$ scattering,
the contribution from the $\Sigma^0$ diagram is less 
than one seventh of the contribution
from the $\Sigma^+$ channel. 
Further, recently
it was suggested that 
for the 
$\phi$ production~\cite{SIB05} 
a strong enhancement of the reaction
amplitude towards threshold might be due to the presence of a crypto exotic
baryon with hidden strangeness. 
Although this observation is not directly applicable
to other isospin channels, (in the $\Sigma^0 p$ system no 
corresponding structure was observed) it might be a hint of an exotic
mechanism. 
In any case,
present theoretical predictions of
the cross sections strongly underestimate
the experimental data. The comparison of the excitation function
expected based upon phase space distribution including $N-Y$ FSI to the
data results in FSI parameters comparable to the $p-\Lambda$ system. 
This may indicate a strong $n-\Sigma^+$ interaction
but due to the large systematic uncertainties the data are also
consistent with a pure phase space distribution without $n-\Sigma^+$ FSI. 

Further studies of the $\Sigma^+$ production are necessary to
clarify the picture. On the experimental side additional data points
should be added for which an improved event selectivity is favorable
to reduce the large uncertainties introduced by the background
subtraction. A $4 \pi$ detection system for  neutral and charged
particles which will be soon available with WASA at COSY could be
used~\cite{WASA}. On the theoretical side an improved model has to be
developed which consistently reproduces the hyperon cross section data
near threshold.

\section{Acknowledgments}

        \label{acknow}

This work has been supported by the European Community - Access to
Research Infrastructure action of the Improving Human Potential
Programme, by the FFE grants (41266606 and 41266654) from the Research
Center J\"ulich, by the DAAD Exchange Programme (PPP-Polen), and by
the Polish State Committee for Scientific Research (grant
No. PB1060/P03/2004/26).

\section{Note added in proof}
  
During the evaluation process of the article we have been made aware
of the predictions of the excitation function of the total cross section
for the \rS reaction which is closer to the data in comparison with 
the models discussed, yet still underpredict the determined total cross sections
by more than an order of magnitude~\cite{shyamnew}.
   


\end{document}